\documentclass[prl,amsmath,amssymb,twocolumn,showpacs,footinbib]{revtex4}

\usepackage{amsmath}
\usepackage{graphicx}
\usepackage{subfigure}
\usepackage{adjustbox}
\usepackage{bm}
\usepackage{color}
\usepackage{braket}
\usepackage{standalone}
\usepackage{multirow}
\usepackage{tikz}
\usepackage{mathrsfs}
\usepackage{dsfont}
\usepackage{amssymb}
\usepackage[colorlinks,bookmarks=true,citecolor=blue,linkcolor=red,urlcolor=blue]{hyperref}

\begin{document}
\title{Non-Hermitian extensions of higher-order topological phases \\ and their biorthogonal bulk-boundary correspondence}

\author{Elisabet Edvardsson, Flore K. Kunst, and Emil J. Bergholtz}

\affiliation{Department of Physics, Stockholm University, AlbaNova University Center, 106 91 Stockholm, Sweden}
\date{\today}

\begin{abstract}
Non-Hermitian Hamiltonians, which describe a wide range of dissipative systems, and higher-order topological phases, which exhibit novel boundary states on corners and hinges, comprise two areas of intense current research. Here we investigate systems where these frontiers merge and formulate a generalized biorthogonal bulk-boundary correspondence, which dictates the appearance of boundary modes at parameter values that are, in general, radically different from those that mark phase transitions in periodic systems. By analyzing the interplay between corner/hinge, edge/surface and bulk degrees of freedom we establish that the non-Hermitian extensions of higher-order topological phases exhibit an even richer phenomenology than their Hermitian counterparts and that this can be understood in a unifying way within our biorthogonal framework. Saliently this works in the presence of the non-Hermitian skin effect, and also naturally encompasses genuinely non-Hermitian phenomena in the absence thereof. 
\end{abstract}

\maketitle

\textit{Introduction.}
Topological phases of matter are at the forefront of condensed-matter research with a recent focus on higher-order topological phases \cite{benalcazarbernevighughes, langhehnpentrifuoppenbrouwer, linhughes, schindlercookvergnio, parameswaranwan, imhofbergerbayerbrehmmolenkampkiesslingschindlerleegreiterneupertthomale, ashraf,benalcazarbernevighughesagain,
garciaperissstrunkbilallarsenvillanuevahuber,petersonbenalcazarhughesbahl,songfangfang,ezawapap,KuMiBe2018}, where a subtle interplay between topology and crystalline symmetry results in the appearance of boundary states on boundaries with a codimension higher than one, i.e., corners or hinges. Another increasingly popular direction of research revolves around studying topology in the context of non-Hermitian physics, which is a relevant approach for describing a wide range of dissipative systems \cite{MaAlVaVaBeFoTo2018,HatanoNelson, Xi2018, KuEdBuBe2018, KuDw2018,lee,yaosongwang,yaowang,leethomale,schomerus,gong,carlstroembergholtz,jan,koziifu,knots,knots2,yoshidapeterskawakmi,NHarc,wiemannkremerplotniklumernoltemakrissegevrechtsmanszameit,NHtransition,NHexp,EPringExp,NHexp2,NHlaser,asymhop1,asymhop2,yuce}. Saliently these models feature a breakdown of the conventional bulk-boundary correspondence \cite{Xi2018, KuEdBuBe2018, KuDw2018,lee,yaosongwang,yaowang,leethomale}, which is intimately linked to the piling up of ``bulk" states at the boundaries known as the \emph{non-Hermitian skin effect} \cite{MaAlVaVaBeFoTo2018,KuEdBuBe2018,yaowang}. These models can be understood with open boundaries directly by defining a \emph{biorthogonal bulk-boundary correspondence} \cite{KuEdBuBe2018}, which combines the right and left wave functions of the boundary modes to each other to form a ``biorthogonal state." By studying the behavior of this state, it is then possible to reconcile the physics of open non-Hermitian systems.

Here we show that the concept of a biorthogonal bulk-boundary correspondence can be generalized to capture non-Hermitian extensions of higher-order topological phases. Indeed, such phases have very recently been studied in a number of works resulting in the observation of variations to the skin effect and the suggestion of topological invariants \cite{LiZhAiGoKaUeNo2018, LeLiGo2018, Ez2018, Ez20182}. Here, unlike Refs.~\onlinecite{LiZhAiGoKaUeNo2018, LeLiGo2018, Ez2018, Ez20182}, we focus on the biorthogonal properties of the open boundary systems, and show that this provides a comprehensive and transparent interpretation of the physical features of non-Hermitian extensions of higher-order topological phases. In particular, it unravels a subtle interplay between crystalline lattice symmetries, sample geometry, and boundary/bulk states that goes qualitatively beyond that of the Hermitian realm.  

To elucidate these results we introduce several pertinent examples that admit an exact analytical treatment. First we investigate a non-Hermitian chiral hinge insulator where the conventional bulk-boundary correspondence is broken: the presence of open boundaries drastically rearranges the entire energy spectrum concomitant with a macroscopic piling up of states at the hinges. Second, we find corner modes on two geometries of the breathing kagome lattice, the rhombus and the triangle. Interestingly, in the case of the rhombus the open boundary conditions lead to the appearance of additional \emph{biorthogonal bulk states} in a regime that is traditionally associated with edge bands. For the triangle geometry, however, no such effect is observed, but instead the corner states disappear to the bulk first via an edge transition, which has no counterpart in the Hermitian limit. In each case we show that these features can be naturally understood at a microscopic level by analyzing the biorthogonal set of exact analytical expressions for the higher-order boundary states. 

\begin{figure*}
	\graphicspath{{FiguresNH/}}
	\includegraphics[width=0.88\linewidth]{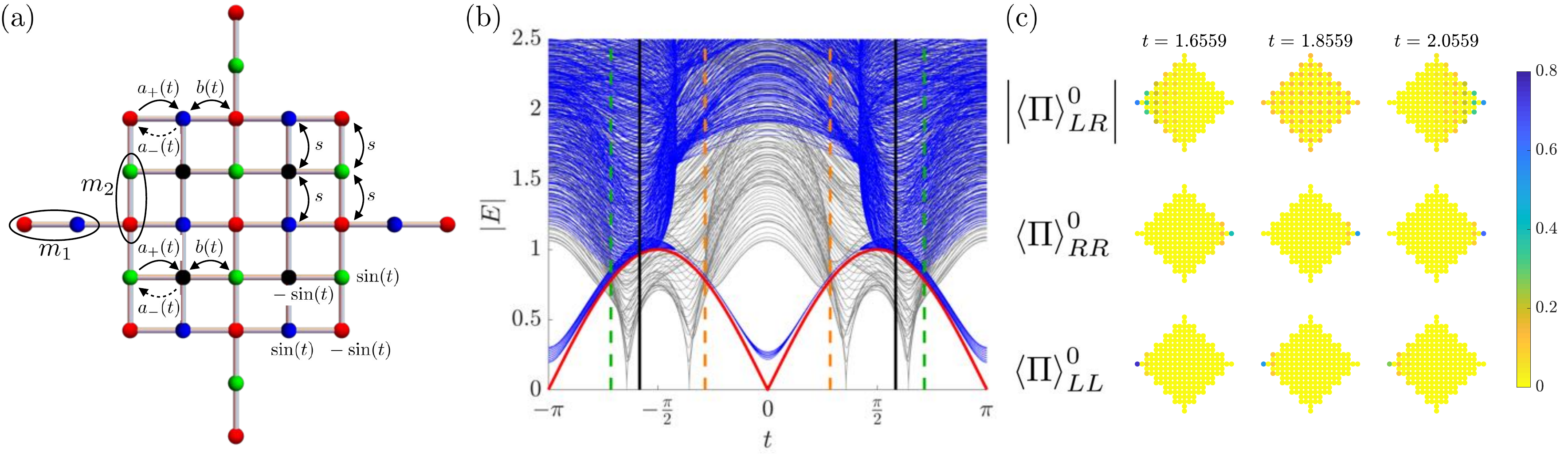}
	\caption{(a) Schematic depiction of the lattice with $A$, $B$, $B'$, and $B''$ sublattices in red, blue, green and black, respectively, and the unit cells labeled by $m_1$ and $m_2$. The red-blue and green-black chains are coupled to each other via nearest-neighbor hopping, $s$, and the Hamiltonian on each of these two types of chains has the onsite term $(-) \textrm{sin}(t)$, and hopping terms $a_\pm (t)  = -t_1 + \delta \,  \textrm{cos}(t) \pm \gamma/2$ and $b(t) = -t_1 - \delta \, \textrm{cos}(t)$, where $t_1$, $\gamma$, and $\delta$ are nearest-neighbor hopping parameters. (b) The absolute value of the energy with $\gamma= 2.5$, $t_1 = 1.5$, $\delta= 1$, $s = 0.25$, and $M_1 = M_2 = 11$ as a function of $t$ is shown in blue (gray) for the open (periodic) system. The orange (green) dashed lines correspond to $|r_{R,1}| = 1$ ($|r_{L,1}| = 1$), while the black solid lines correspond to $|r_{L,1} r_{R,1}| = 1$. (c) The localization of the hinge state [in red in (b)] on a lattice with $M_1 = M_2 = 11$ computed using the biorthogonal (top row), right (middle), and left (bottom) expectation values of the projection operator.}
	\label{fig:hinge_model}
\end{figure*}

\textit{Non-Hermitian chiral hinge insulator.} We start by studying the lattice in Fig.~\ref{fig:hinge_model}(a) with open boundaries in two directions while being periodic in the third dimension parametrized by $t$. Each red-blue and green-black chain represents a one-dimensional charge pump, and the Hamiltonian of each of these chains, explicitly shown in Fig.~\ref{fig:hinge_model}(a), corresponds to the Rice-Mele model in the Hermitian limit \cite{RiMe1982}, such that these chains individually realize a Chern insulator with opposite Chern number on the differently colored chains. The Hermitian limit of this model is known to have exactly solvable temporal chiral hinge states, which are protected by a mirror Chern number \cite{KuMiBe2018, KuMiBe20182}.

We implement non-Hermiticity by introducing a preferred hopping direction between unit cells on the individual red-blue and green-black chains. Explicitly, we change the magnitude of hopping to the right with respect to hopping to the left yielding a nonreciprocal tight-binding model \cite{footnote2}. The chains are then coupled to each other in a Hermitian fashion [cf. Fig.~\ref{fig:hinge_model}(a)], such that the non-Hermiticity only presides in one direction.

The absolute value of the band spectrum for this model with open boundary conditions is shown in Fig.~\ref{fig:hinge_model}(b) \cite{footnote}, where the chiral hinge state is shown in red, the bulk bands in blue, and the bunched blue bands appearing in the gap are traditionally identified as surface bands. While the open spectrum is that of a chiral hinge insulator, the periodic Bloch spectrum indicated in gray is semimetallic, thus manifesting a striking breakdown of conventional bulk-boundary correspondence. 

\begin{figure}[b]
	\graphicspath{{FiguresNH/}}
	\includegraphics[width=0.88\linewidth]{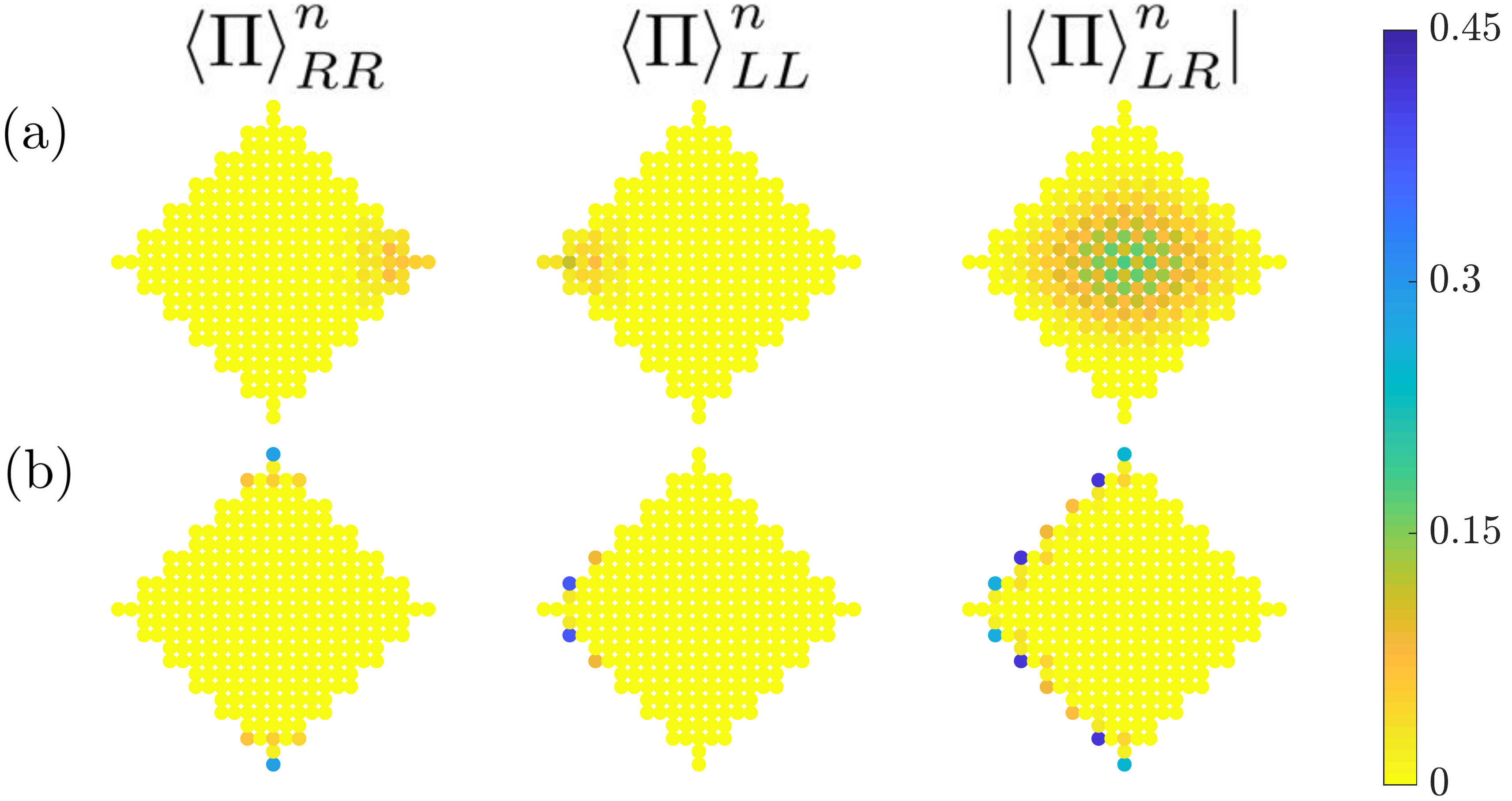}
	\caption{Distribution of the (a) bulk and (b) surface band with energies (a) $|E| = 2.0264$ and (b) $|E| = 0.5345$ for the same parameters as in Fig.~\ref{fig:hinge_model}(b) for the cut $t=-0.5$. The left and middle columns show the localization of the right and left wave function individually, and the right column shows the biorthogonal localization.}
	\label{fig:hinge_skin_effect}
\end{figure}

\begin{figure*}[t]
	\graphicspath{{FiguresNH/}}
	\includegraphics[width=0.88\linewidth]{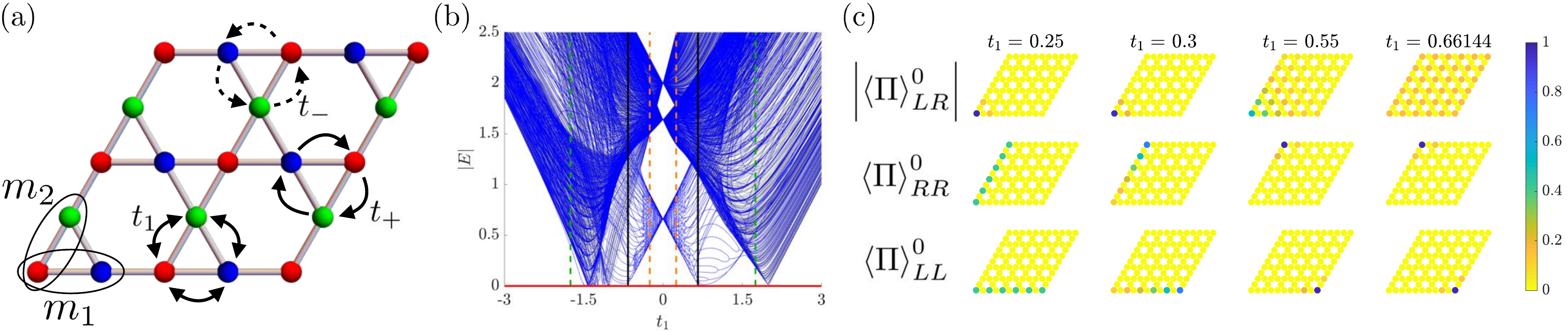}
	\caption{(a) Schematic depiction of the lattice with $A$, $B$, and $B'$ sublattices in red, blue, and green, respectively, and the unit cells labeled by $m_1$ and $m_2$. In the Hamiltonian the sites in the ``up" triangles are connected via a nearest-neighbor hopping $t_1$, while in the ``down" triangles, the nearest-neighbor hopping in a clockwise fashion is $t_+ = t_2 + \gamma/2 $, and $t_- = t_2 - \gamma/2 $ in the counterclockwise motion. (b) The absolute value of the energy with $t_2 = 1$, $\gamma= 1.5$, and $M_1 = M_2 = 20$ as a function of $t_1$ is shown in blue (gray) for the open (periodic) system. The orange (green) dashed lines correspond to $|r_{R,1}| = |r_{L,2}| = 1$ ($|r_{R,2}| = |r_{L,1}| = 1$), while the black solid lines correspond to $|r_{L,1} r_{R,1}| = |r_{L,2} r_{R,2}| = 1$. (c) The localization of the corner state [in red in (b)] on a lattice with $M_1 = M_2 = 6$ computed using the biorthogonal (top row), right (middle), and left (bottom) expectation values of the projection operator.
	}
	\label{fig:kagome_rhombus}
\end{figure*}

To rationalize this behavior we study the distribution of bulk bands $E_n$ in the lattice at a specific cut of $t$ in Fig.~\ref{fig:hinge_model}(b) by computing $\left<\Pi_{m_1,m_2}\right>_{\alpha \alpha'}^n\equiv\braket{\Psi_{\alpha,n}|\Pi_{m_1,m_2}|\Psi_{\alpha',n}}$ for $\alpha, \alpha' \in \{R,L\}$, where $\Pi_{m_1,m_2} = \sum_{\beta \in \{A,B,B',B''\}} \ket{e_{\beta,m_1,m_2}}\bra{e_{\beta,m_1,m_2}}$ is a projection operator onto each site in unit cell $\{m_1,m_2\}$ and $\ket{\Psi_{R,n}}$ ($\ket{\Psi_{L,n}}$) the associated right (left) wave function. The quantities  $\left<\Pi_{m_1,m_2}\right>_{RR}^n$ and $\left<\Pi_{m_1,m_2}\right>_{LL}^n$ are similar to what is known in ordinary quantum mechanics as the expectation value, and we show them in the left and middle panel of Fig.~\ref{fig:hinge_skin_effect}(a), respectively. We see that the bulk state is localized to the right and left hinge, respectively, such that the breaking of bulk-boundary correspondence indeed goes hand in hand with the piling up of states, as was also observed in Refs.~\onlinecite{LiZhAiGoKaUeNo2018, LeLiGo2018, Ez20182}. Interestingly, if we now consider the \emph{biorthogonal expectation value} $\left<\Pi_{m_1,m_2}\right>_{LR}^n$, we find the distribution displayed in the right panel of Fig.~\ref{fig:hinge_skin_effect}(a), which is in accordance with expected bulk-band behavior. Therefore, we label the blue bands \emph{biorthogonal bulk bands}. Similarly, when studying the localization of a surface band, i.e., a band that belongs to the bunched blue bands in the band gap, we find that the right wave function is localized to the top and bottom hinge [cf. left panel of Fig.~\ref{fig:hinge_skin_effect}(b)] while the left wave function lives on the left hinge [cf. middle panel of Fig.~\ref{fig:hinge_skin_effect}(b)]. The biorthogonal expectation value, however, reveals that the weight of the state is indeed distributed on the surfaces [cf. right panel of Fig.~\ref{fig:hinge_skin_effect}(b)], and we call this a \emph{biorthogonal surface state}. While we thus observe anomalous ``skin" behavior when investigating the spectrum with reference to its right (or left) wave functions only, we observe an ``ordinary" distribution of the bulk and surface states when approaching the problem from a biorthogonal perspective. These observations are natural considering that the relation between the eigenstates, the Hamiltonian, and the energies involves the left {\it and} right eigenstates as $E_n=\left<H\right>_{LR}^n$.

Next we turn to the hinge state [red in Fig.~\ref{fig:hinge_model}(b)], and generalize Refs.~\onlinecite{KuMiBe2018, KuMiBe20182} to write down the exact solutions 
\begin{equation}
\ket{\psi_\alpha} = \mathcal{N}_\alpha\sum_{m_1,m_2}^{}\left(r_{\alpha,1}\right)^{m_1}\left(r_{\alpha,2}\right)^{m_2}c^{\dagger}_{A,m_1,m_2}\ket{0}, \label{eq:exact_sol}
\end{equation}
which have the remarkable property that they may localize on opposite hinges depending on $\alpha\,{\in}\,\{R, L\}$ which labels the right and left eigenvectors. Here $m_1$ and $m_2$ label the unit cells in the lattice with a total of $M_1M_2$ unit cells, $\mathcal{N}_\alpha$ is the normalization constant, $c^{\dagger}_{A,m_1,m_2}$ creates a particle on the $A$ sublattice [in red in Fig.~\ref{fig:hinge_model}(a)] in unit cell $\{m_1,m_2\}$, and $r_{\alpha,1}$ and $r_{\alpha,2}$ can be computed analytically and read
$r_{R,1}=-\frac{-t_1+\cos(t)-\gamma/2}{-t_1-\cos(t)}, 
r_{L,1} =-\frac{-t_1+\cos(t)+\gamma/2}{-t_1-\cos(t)},
r_{R,2} = r_{L,2} = -1.$ These wave functions have zero amplitude on all blue, green, and black sites, and the associated eigenenergy corresponds to the eigenenergy on the $A$ sublattice, which is $E_0\,{=}\, {-}\,\textrm{sin}(t)$ in accordance with the red band in Fig.~\ref{fig:hinge_model}(b). Depending on the values of $|r_{R,1}|$ and $|r_{L,1}|$, the state $\ket{\psi_\alpha}$ behaves as a hinge or a bulk state \cite{footnote3}. In particular, the right (left) wave function is equally localized on each $A$ sublattice, in which case it behaves as a bulk state, when $|r_{\alpha,1}| \,{=}\,1$ corresponding to the orange and green dashed lines in Fig.~\ref{fig:hinge_model}(b). This, however, is in disagreement with the attachment of the red band to the bulk bands, where there is indeed a small gap between the red band and the blue surface bands at the orange lines.

Instead, we consider $|r_{L,1} r_{R,1}| = |r_{L,2} r_{R,2}| = 1$, and find an accurate prediction for hinge-state attachment to the bulk [cf. the black solid lines in Fig.~\ref{fig:hinge_model}(b)]. This quantity follows from considering the biorthogonal expectation value of the projection operator $\left<\Pi_{m_1,m_2}\right>_{LR}^0$ using the solutions in Eq.~(\ref{eq:exact_sol}) \cite{KuEdBuBe2018}, which is plotted in the top row of Fig.~\ref{fig:hinge_model}(c) for three cuts in $t$. Indeed, we see that while the biorthogonal product indicates bulk-band behavior in the middle column, the right and left wave functions shown in the middle and bottom rows of Fig.~\ref{fig:hinge_model}(c), respectively, suggest the band is localized to the hinge. We thus find that to accurately describe the physics of a non-Hermitian system with open boundary conditions, one has to invoke a biorthogonal bulk-boundary correspondence \cite{KuEdBuBe2018}. Moreover, we find that the hinge state only changes localization upon attachment to the bulk bands in full agreement with the Hermitian version of this model \cite{KuMiBe2018}.

\begin{figure*}[t]
	\graphicspath{{FiguresNH/}}
	\includegraphics[width=0.88\linewidth]{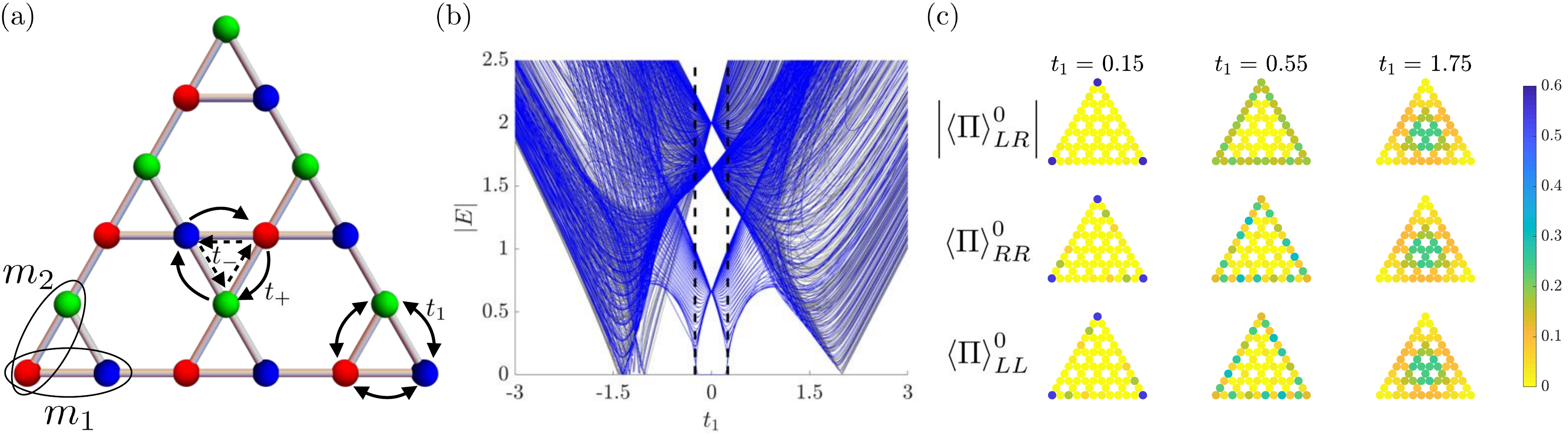}
	\caption{(a) Schematic depiction of the lattice with $A$, $B$, and $B'$ sublattices in red, blue and green, respectively, and the unit cells labeled by $m_1$ and $m_2$. The Hamiltonian is explicitly depicted, and the same as in Fig.~\ref{fig:kagome_rhombus}(a). (b) The absolute value of the energy with $t_2=1$, $\gamma= 1.5$, and $M_1 = M_2 = 20$ as a function of $t_1$ is shown in blue (gray) for the open (periodic) system. The black dashed lines correspond to $|r_{R,2}| = |r_{L,1}| = 1$. (c) The localization of the lowest energy band on a lattice with $M_1 = M_2 = 6$ for different cuts in $t_1$ using the biorthogonal (top row), right (middle), and left (bottom) expectation values of the projection operator.}
	\label{fig:kagome_triangle}
\end{figure*}

\textit{Breathing kagome lattice.} Next we study the two-dimensional, breathing kagome lattice in the geometry of a rhombus [cf. Fig.~\ref{fig:kagome_rhombus}(a)] and a triangle [cf. Fig.~\ref{fig:kagome_triangle}(a)]. We implement non-Hermiticity in these lattices by changing the magnitude of the hopping terms in the down triangles such that the hopping amplitude in the clockwise direction, $t_+= t_2 + \gamma/2$, is unequal to the hopping in the anticlockwise direction, $t_- = t_2 - \gamma/2$ while keeping the hopping on up triangles, $t_1$, nonchiral. The real-space Hamiltonian for both models is explicitly shown in Figs.~\ref{fig:kagome_rhombus}(a) and \ref{fig:kagome_triangle}(a), respectively. The Hermitian versions of these systems were previously studied in Refs.~\onlinecite{ezawapap,KuMiBe2018, xuxuewan,akari}.

We start by focusing on the rhombus, and plot the absolute value of the energy spectrum in Fig.~\ref{fig:kagome_rhombus}(b) as a function of $t_1$ for fixed $t_2$ and $\gamma$ with the bulk bands in blue, and the red band corresponding to a zero-energy corner mode \cite{footnote}. In addition, we plot the spectrum with periodic boundary conditions in gray, and find that it is qualitatively in accordance with the spectrum in blue. This can be understood from the fact the bulk states can move around in loops in the lattice, such that they do not get trapped, and thus do not pile up. Nevertheless, when considering the behavior of the blue bands that do not overlap with the gray bands, which are thus expected to be associated with surface states, we find that some of these states are in fact \emph{additional} biorthogonal bulk states \cite{footnote2}. 

To study the behavior of the zero-energy corner mode in more detail, we consider its associated exact wave-function solution given in Eq.~(\ref{eq:exact_sol}) with
\begin{align}
r_{R,1} &= r_{L,2} = - \frac{t_1}{t_2 {+} \gamma/2}, \quad r_{R,2} = r_{L,1} = - \frac{t_1}{t_2 {-} \gamma/2}. \label{eq:sol_kagome_rhombus}
\end{align}
Depending on the values of $|r_{\alpha,1}|$ and $|r_{\alpha,2}|$ with $\alpha \in \{R,L\}$, the state $\ket{\Psi_\alpha}$ thus behaves as a corner, edge or bulk state. Considering $|r_{R,1}| = |r_{R,2}|=1$ and $|r_{L,1}| = |r_{L,2}|=1$, which would predict a bulk state in the framework of ordinary quantum mechanics, we find solutions corresponding to the orange and green dashed lines, respectively, in Fig.~\ref{fig:kagome_rhombus}(b), and clearly see that this does not predict any particular behavior in the spectrum. Instead, considering the biorthogonal delocalization criteria for $\left<\Pi_{m_1,m_2}\right>_{LR}^0$,   corresponding to $|r_{L,1} r_{R,1}| = |r_{L,2} r_{R,2}| = 1$, we obtain the black solid lines in Fig.~\ref{fig:kagome_rhombus}(b), in complete agreement with attachment to the bulk bands.

To corroborate this picture we plot $\left<\Pi_{m_1,m_2}\right>_{\alpha \alpha'}^0$ with $\alpha, \alpha' \in \{R,L\}$ in Fig.~\ref{fig:kagome_rhombus}(c) for four different choices of $t_1$. This  reveals several aspects. First of all, the right and left wave functions (middle and bottom row, respectively) individually suggest a transition of the corner state from one corner to the other via the \emph{edges}, while the biorthogonal state (top row) reveals a transition via the \emph{bulk}. Indeed, only the latter interpretation is in accordance with the Hermitian version of this model \cite{KuMiBe2018, ezawapap, xuxuewan}. Secondly, we study the blue bands to which the corner state attaches at $|r_{L,1} r_{R,1}| = |r_{L,2} r_{R,2}| = 1$, and find that they are edge bands in the context of their right and left wave functions, while admitting a bulk channel when considering their biorthogonal properties \cite{footnote2}, i.e., they correspond to the aforementioned biorthogonal bulk states. This thus explains the migration of the biorthogonal corner state into the bulk, as opposed to it being transmitted via the edges in the case of the right and left wave functions. Third, we observe that the right (left) wave function of the corner state localizes to the top left (bottom right) corner for $t=0.55$ [cf. third column from the left in Fig.~\ref{fig:kagome_rhombus}(c)], while the corner mode is well separated from the other bands in the spectrum. This is noteworthy because in the Hermitian limit, localization of the corner state to these specific corners is not possible \cite{KuMiBe2018, xuxuewan}.


Next, we turn to the triangular geometry in Fig.~\ref{fig:kagome_triangle}(a). We plot the absolute value of the band spectrum in Fig.~\ref{fig:kagome_triangle}(b) with the bands for open (periodic) boundary conditions in blue (gray) \cite{footnote}, and again find that the bulk spectrum is not qualitatively rearranged signaling the absence of the skin effect. As in the Hermitian case \cite{ezawapap}, there is a threefold degenerate zero-energy mode for a certain parameter regime. To understand the behavior of this mode in more detail, we note that the solutions in Eq.~(\ref{eq:exact_sol}) with $r_{\alpha,1}$ and $r_{\alpha,2}$ given in Eq.~(\ref{eq:sol_kagome_rhombus}) in the large system limit can be mapped onto each corner of the triangle for $|r_{\alpha,1}|,|r_{\alpha,2}| <1$. Once $|r_{\alpha,1}|$ and/or $|r_{\alpha,2}|$ become(s) equal to or larger than $1$, the states $\ket{\psi_\alpha}$ leak into the edge or the bulk such that the three corner states start to interfere with each other. This renders the predictive power of Eq.~(\ref{eq:exact_sol}) invalid, while also resulting in the lifting of the zero-energy mode away from zero, which, assuming $t_1, t_2, \gamma \in \mathbb{R}$, indeed happens once $|t_1| = |t_2 - \gamma/2|$, i.e., $|r_{R,2}| = |r_{L,1}| = 1$ [cf. the black dashed lines in Fig.~\ref{fig:kagome_triangle}(b)]. Studying the localization of the lowest-lying energy mode for different choices of $t_1$ in Fig.~\ref{fig:kagome_triangle}(c), we see that the right, left, and biorthogonal distributions predict the same qualitative behavior \cite{footnote2}. Indeed, in the left column we consider $t_1=0.15$ for which the lowest-energy mode is the zero-energy corner state, and see that they are equally distributed over the three corners. When investigating the distribution of the states when they attach to the edge bands [cf. black dashed lines in Fig.~\ref{fig:kagome_triangle}(b)], we indeed see edge band behavior, while attachment to the bulk bands results in bulk-band behavior [cf. right column in Fig.~\ref{fig:kagome_triangle}(c)]. In sharp contrast to the Hermitian version of this model \cite{ezawapap}, we find that there is a corner to edge to bulk transition, which confirms the prediction following from Eqs.~(\ref{eq:exact_sol}) and (\ref{eq:sol_kagome_rhombus}) that the corner mode first attaches to the edge states before merging with the bulk bands. This behavior can be understood from symmetry arguments---the threefold rotation dictates the same behavior for all three corner states, which due to the presence of non-Hermitian terms necessarily leak into the edge \emph{before} they enter the bulk.

\textit{Conclusion}. We have considered three explicit examples of non-Hermitian extensions of second-order topological systems, and shown that while conventional bulk-boundary correspondence may be strongly broken, we can exploit the biorthogonal properties of these models to fully reconcile their behavior also in the presence of a skin effect. By making use of exact solutions for the second-order boundary states [cf. Eq.~(\ref{eq:exact_sol})], we explicitly studied the localization of these states both in the context of their right and left wave functions as well as their biorthogonal product, and related that to the spectrum with open boundary conditions. Moreover, by studying the distribution of edge/surface and bulk bands of these models we showed that additional biorthogonal bulk bands may appear when taking open boundary conditions, and that the interplay between edge, corner, and bulk can be qualitatively distinct from that of their Hermitian counterparts even in the absence of the non-Hermitian skin effect.

\acknowledgments
{\it Acknowledgments.} 
We thank Jan Carl Budich and Guido van Miert for related collaborations. E.E. thanks Eva Mossberg for useful discussions about MATLAB. This work was supported by the Swedish Research Council (VR) and the Wallenberg Academy Fellows program of the Knut and Alice Wallenberg Foundation.


\begin{thebibliography}{10}
\bibitem{footnote3}
Note that if $|r_{R,2}|, |r_{L,2}| \neq 1$, $r_{\alpha,1}$ and $r_{\alpha,2}$ with $\alpha \in \{R,L\}$ could be chosen such that the wave function in Eq.~(\ref{eq:exact_sol}) corresponds to a surface state.
\end{thebibliography}
\end{document}